# PUB-MS – a mass-spectrometry–based method to monitor protein-protein proximity *in vivo*.

Arman Kulyyassov[1,2], Muhammad Shoaib[1], Andrei Pichugin[1], Patricia Kannouche[1], Erlan Ramanculov[2], Marc Lipinski[1] and Vasily Ogryzko[1*]

1 - Institut Gustave Roussy, 39 Rue Camilles Desmoulin, 94805, Villejuif, France

2 - National Center of Biotechnology, Valihanova 43, 01000, Astana, Kazakhstan

\* - corresponding author;

- e-mail: vogryzko@gmail.com
- phone : 33 – 1 42 11 65 25
- FAX :   33 – 1 42 11 65 25

**Running title:** Proximity-Utilizing Biotinylation

**Abbreviations:** BAP – Biotin Acceptor Peptide; LC-MS/MS – liquid chromatography coupled with tandem mass-spectrometry; TIC – Total Ion Current

# Abstract

The common techniques to study protein-protein proximity *in vivo* are not well-adapted to the capabilities and the expertise of a standard proteomics laboratory, typically based on the use of mass spectrometry. With the aim of closing this gap, we have developed PUB-MS (for Proximity Utilizing Biotinylation and Mass Spectrometry), an approach to monitor protein-protein proximity, based on biotinylation of a protein fused to a biotin-acceptor peptide (BAP) by a biotin-ligase, BirA, fused to its interaction partner. The biotinylation status of the BAP can be further detected by either Western analysis or mass spectrometry. The BAP sequence was redesigned for easy monitoring of the biotinylation status by LC-MS/MS. In several experimental models, we demonstrate that the biotinylation *in vivo* is specifically enhanced when the BAP- and BirA- fused proteins are in proximity to each other. The advantage of mass spectrometry is demonstrated by using BAPs with different sequences in a single experiment (allowing multiplex analysis) and by the use of stable isotopes. Finally, we show that our methodology can be also used to study a specific subfraction of a protein of interest that was in proximity with another protein at a predefined time before the analysis.

**Keywords**: Protein-protein interactions, biotinylation, proximity, multiplexing, pulse-chase, SILAC

# Introduction

One of the ultimate goals of molecular biology is reconstruction of spatio-temporal structure of a living cell at the molecular level. This task includes determination (and cataloguing) of proximities between different molecular components in the cell and monitoring their time- and physiological state- dependent changes. In many cases, proximity between macromolecules arises due to their interactions (either direct or else via intermediates), however, the contribution of dynamic self-organization in generation of spatio-temporal order is emerging as another viable possibility [1, 2]. Specifically, in proteomics, this implies that the detection of protein-protein *proximity* is a more general task than gaining information about *physical interactions* between proteins, as it could detail aspects of spatial order *in vivo* that are challenging (at the very least) to reconstitute in binding experiments *in vitro*.

In recent years, several approaches have been developed to monitor protein-protein interactions inside living mammalian cells, and thus to provide the physiologically relevant information about the spatial organization of cell. They include Fluorescence Resonance Energy Transfer (FRET) with fluorescent proteins [3, 4], Bioluminescence Resonance Energy Transfer (BRET) [5, 6], Protein Complementation Assays (PCA) [7, 8], and various two-hybrid systems [9, 10]. However, they suffer from several limitations that we aimed to overcome in our work. For example, only relatively close proximity can be studied with these methods (for example, FRET provides intra- or intermolecular distance data in the range 1-10 nm). Second, although the study of protein-protein interactions falls within the scope of proteomics, the FRET/BRET and PCA methods require access to instrumentation (e.g., confocal microscope) that usually do not match the capabilities and the expertise of a standard proteomics laboratory.

Given that a mass spectrometer is the working horse of a core proteomics facility, it is desirable to develop a mass-spectrometry-oriented approach to monitor protein-protein interactions (and, more generally, their proximity) *in vivo*. Since accurate mass measurement is the main principle of this methodology, such an approach could be based on the proximity-dependent introduction of a modification in the tested protein, which can be further detected in a quantitative manner with the help of mass spectrometry. One potential advantage of mass-spectrometry, as compared to other methods (such as FRET, BRET, PCA, two-hybrid) is the high resolution of mass spectrometer, i.e., its ability to reliably distinguish between thousands of ions with different m/z ratios. This opens a possibility of multiplexing, i.e., monitoring of several interactions in a single experiment.

Previously, we and others have developed a method for efficient biotinylation of proteins *in vivo* [11, 12], based on the coexpression within the same cell, of the biotin ligase BirA and the protein of interest fused to a short biotin acceptor peptide (BAP). Such *in vivo* biotinylated proteins have been used for protein complex purification [11, 13, 14], chromatin immunoprecipitation [12, 13, 15, 16], immunoelectron microscopy [17] and other purposes. In this work, we have adapted our system to the task of monitoring protein-protein proximity *in vivo,* with the use of mass spectrometry. For this purpose, the BirA was fused to one of the interaction partners, whereas the BAP was modified to make the detection of its biotinylation possible by mass spectrometry. Using several experimental systems, we show that the biotinylation of BAP is interaction (and/or proximity) dependent. In addition, we demonstrate that BAP domains with different primary amino acid structures and thus with different molecular weights can be used in the same experiment, providing the possibility of multiplexing. Alternatively to the changes in primary amino acid structure, the stable isotope format can also be used, providing another way to perform multiplexing experiments.

An additional advantage of this technology resides in the fact that it is not limited to detection of short-range proximities, as FRET and PCA are. In particular, we demonstrate that sharing the same intracellular compartment can also give a signal above the background. Thus, with appropriate controls, our approach can provide a new kind of information, complementary to the short-range proximity data obtained by the FRET and PCA methodologies.

Finally, we also demonstrate that our system can help to overcome another limitation of current methodologies to detect protein-protein proximity. Given an introduction of a covalent mark in the protein of interest, one can ask questions that are impossible to address with FRET/BRET or PCA. For example, one can follow the state of a protein of interest at a defined time after its interaction with another protein has occurred. This application should be particularly useful for studying multistep intracellular processes, where the proximities between proteins (e.g., interaction partners) and protein properties typically change in a sequential manner. Thus, in addition to helping to reconstruct the cell topology in space, this approach has a promise to add the temporal dimension to such studies.

## Experimental procedures

*Recombinant DNA*

For the pCMV enhancer vectors, the pCDNA3.1 (Invitrogen) backbone was used. For the MoMuLV enhancer vectors, the pOZ vector was used [18]. The BirA ORF was PCR amplified from the vector pBBHN [12]. The BAP domain fragments were prepared by designing oligonucleotides and sequential PCR, followed by restriction digest and insertion into the pCDNA3.1 vector. The insert sequences were confirmed by sequencing. The vectors generated were used to clone the ORFs of H2AZ, HP1γ, TAP54α, KAP1-HP1BD, Rad18, GFP, H2ABBD, macroH2A and H3.3.

*Cell culture and cell lines*

HEK293T cells were grown in Dulbecco's modified Eagle's medium with high glucose (PAA) and 10% fetal bovine serum (FCS, PAA). For transient transfection and Western analysis, a standard calcium phosphate precipitation method was used, and the cells were analyzed one or two days after transfection, as indicated. For the biotin labeling *in vivo*, cells were grown for several days before transfection in the DMEM supplied with dialyzed FCS, and for the specified time of labeling, biotin (Sigma) was added to a final concentration of 5 μg/ml, while the pH was stabilized by addition of 50 mM HEPES (pH 7.35) to the medium. For SILAC experiments, the cells were grown in DMEM with $^{12}C_6$ L-lysine or $^{13}C_6$ L-lysine (Thermo Scientific, cat # 89983) for at least 5 divisions before transfection, and kept in the same medium until harvest. For the Western analysis, $3*10^5$ of cells (corresponding to one well in a 6-well plate) were used for transfection. For the mass-spectrometry analysis, $3*10^6$ of cells were transfected per data point.

*Biochemistry and Western blot analysis*

Except where indicated otherwise, cell nuclei were used for analysis. They were prepared by cell disruption in CSK buffer (100 mM NaCl, 300 mM Sucrose, 10 mM Tris pH 7.5, 3 mM $MgCl_2$, 1 mM EGTA, 1.2 mM PMSF, 0.5% Triton X-100), and centrifugation for 5 min at 4000 rpm. For the analysis of chromatin-associated histones (chromatin fraction), the nuclei were first incubated in CSK buffer containing 450 mM NaCl by 30 min rotation at 4° C, then spun at 4000 rpm, and the supernatant containing soluble histones was discarded. For Western analysis, 1x NuPAGE LDS Sample buffer (Invitrogen) with DTT (10 mM) was added, the nuclei were sonicated, boiled for 5 min at 96° C and loaded on 4-12% gradient Novex Tris-Glycine precast gels. After separation, the proteins were transferred to nitrocellulose membranes and probed with HRP-conjugated streptavidin (Sigma, # S5512) or HRP conjugated α-PentaHis antibody (QiaGen, # 34460) according to the manufacturer's protocol, except that for the detection of biotinylated proteins by the HRP-conjugated streptavidin, 500 mM NaCl was added to the washing buffer (PBS + 0.1% Tween). For protein visualization, the gels were stained with PageBlue (Fermentas, # R0579). For the densitometric analysis of Western blots, the program ImageJ 1.42q (freely available online) was used. To compare the biotinylation levels between different samples, the value of the streptavidin-HRP signal was normalized by taking a ratio with the value of α-His signal, which reflects the amount of the transfected protein regardless of its biotinylation status. Every Western analysis was performed three times, with a representative figure shown.

For $Ni^{2+}$NTA purification of 6XHis tagged proteins, the nuclei were solubilized in buffer A (10% glycerol, 250 mM NaCl, 6 M Guanidine-HCl, 20 mM TrisHCl (pH 8.0), 0.1% Tween) by rotation for 30 min at 4° C. 1/10 of the volume of $Ni^{2+}$NTA agarose, prewashed in the same buffer, was added to the lysate, followed by rotation at 4° C for 3 hr. The beads with bound proteins were washed twice with buffer A, and then twice with buffer B (10% glycerol, 250 mM NaCl, 20 mM TrisHCl (pH 8.0), 0.1% Tween, 0.2 mM PMSF and protease inhibitor cocktail complete (Roche)). The bound proteins were eluted by incubating the beads in 4 volumes of buffer C (10% glycerol, 250 mM NaCl, 20 mM Tris HCl (pH 8.0), 0.1% Tween, 0.2 mM PMSF, 300 mM Imidazole, 50 mM EDTA) and concentrating the sample by ultrafiltration with Microcon YM-10 (Millipore). Alternatively to elution, the beads were washed twice with 25 mM ammonium bicarbonate and direct on-beads digestion was performed.

*Mass spectrometry analysis*

The protein bands were excised from the gel and processed as in [19]. The gel slices were dehydrated with 300 µl of 50% acetonitrile followed by 300 µl of 100% acetonitrile, then re-hydrated with 300 µl of 50 mM ammonium bicarbonate. A final dehydration was performed with 2 washes of 300 µl of 50% acetonitrile, followed by 2 washes of 300 µl of 100% acetonitrile. Each wash was carried out for 10 min at 25° C with shaking at 1400 rpm. The gel slices were dried in a SpeedVac at 35° C for 10 min. For trypsin digestion, the gel slices were pre-incubated with 7 ml of 15 ng/ml trypsin (Promega # V5280) at room temperature for 10 min. Afterwards, 25 µl of 50 mM ammonium bicarbonate was added, and the gel slices were incubated at 37° C for 16 h. The peptide-containing supernatants were dried at 56° C by SpeedVac for 30 min, then resuspended in 20 µl of solution containing 0.05% formic acid and 3% acetonitrile for mass spectrometry experiments. Alternatively, for the on-beads digestion, ammonium bicarbonate was added to the beads at a final concentration of 25 mM, trypsin digestion was performed overnight (12.5 ng/ml), and the peptide mixture was further purified on a Zip tip (Millipore # ZTC18S096). The resulting peptides were analyzed with a nano-HPLC (Agilent Technologies 1200) directly coupled to an ion-trap mass spectrometer (Bruker 6300 series) equipped with a nano-electrospray source. The separation gradient was 7 min from 5% to 90% acetonitrile. The fragmentation voltage was 1.3 V. The analysis of the spectra was performed with the DataAnalysis for the 6300 Series Ion Trap LC/MS Version 3.4 software package. The samples were run in two different modes. For peptide identification, the ion trap acquired successive sets of 4 scan modes consisting of: full scan MS over the ranges of 200-2000 m/z, followed by 3 data-dependent MS/MS scans on the 3 most abundant ions in the full scan. Alternatively, for the confirmation and quantification of the presence of a particular peptide in the sample, the ion trap was set in MRM mode. The sample was separated using the same nanoLC gradient, and the ion trap was set to isolate, fragment and MS/MS scan 5 parental ions having predetermined M/Z ratios. The relative quantity of each peptide in the different fractions was estimated by comparison between the peak areas in the Total Ion Chromatograms (TIC) for this peptide obtained from MRM analysis of these fractions.

*Microscopy*

MRC-5 foetal lung fibroblasts were grown on cover slips in 6-well plates in DMEM high glucose containing dialyzed FBS. Cells were co-transfected with the plasmid constructs BirA-Rad18 (1 microgram) and BAP-H2AZ (1 microgram) using TurboFect™ (Cat. #R0531, Fermentas). 48 h after transfection, cells were treated with UVC (20 J/m$^2$, Philips TUV 15W/G 15 TB lamp). After 6 h of irradiation, the DMEM containing dialyzed FBS was removed and the cells were pulse-labeled for 5 min with biotin (5 μg/mL final concentration). The biotin-containing medium was then removed, and cells were fixed either immediately (pulse) or 2 h after the pulse (chase). For staining, cells were washed once with cold PBS and treated with CSK buffer (100 mM NACl, 300 mM Sucrose, 3 mM MgCl$_2$, 10 mM Tris-HCl pH 7.5, 1 mM EGTA, Triton 0.2%) for 5 min with light agitation. Cells were then washed twice with cold PBS. Fixation was done in 4% formaldehyde for 20 min at room temperature (RT). After fixation, cells were rinsed with ice-cold methanol at -20° C for 10 sec. Cells were then washed 3 times for 5 min with PBS (1X) at RT. Blocking was done for 30 min using 3% BSA and 0.5% Tween20 in PBS. Cells were then incubated with either mouse monoclonal anti-PCNA antibody (Santa Cruz; PC10; sc-56) for 1 h (1/500 in blocking solution) or mouse monoclonal anti-Rad18 antibody (Abcam; ab57447) followed by 3 washes in PBS (1X) at RT. Cells were then incubated with goat anti- mouse secondary antibody conjugated with Alexa-488 (Invitrogen # A11017) for 1 h (1/1000). Streptavidin-Cy3 conjugate (Sigma; # S-6402) was also added to the same incubation mix (1/500). After 3 washes in PBS (1X) at RT, cover slips were mounted on glass slides using VectaShield mounting medium (Vector Laboratories; # H-1000). The cells were observed under a Zeiss LSM 510 Meta confocal microscope, using a Plan-Apochromat 63x 1.4 oil immersion objective. Imaging was performed with sequential multitrack scanning using the 488 and 543 nm wavelengths lasers separately. The colocalization analysis was performed with LSM Examiner software. Pearson's correlation coefficient was calculated with setting the threshold signal common for all images compared.

# Results

**1. Design and features of the system.**

The principle of the method is presented in Figure 1A. Two proteins to be tested for their proximity *in vivo* are coexpressed in the cells of interest, the first protein being expressed as a BirA fusion, and the second one as a fusion to the Biotin Acceptor Peptide (BAP). More efficient biotinylation of the BAP is expected when the two proteins are in proximity to each other, e.g., when interaction occurs (Fig.1A, bottom), as compared to the background biotinylation arising from random collisions between non-interacting proteins (Fig.1A, top). The biotinylation status of the BAP fusion can be further monitored by Western blot or mass spectrometry.

To implement this principle, we constructed two types of vectors: for expression of the BirA fusion and for expression of the BAP fusion, correspondingly (Figure 1B, top). We constructed each vector in two forms, with the protein expression regulated either by a strong CMV or a weaker MoMuLV enhancer. In this work, we used the MoMuLV enhancer for the expression of BirA fusions, while the BAP fusions were expressed from the vectors with the stronger (CMV) promoter. This setting typically allowed us to achieve an excess of the BAP fusion (biotinylation target) over the BirA fusion (biotinylating enzyme), which is essential for the further quantitative analysis.

In addition, we significantly modified the sequence of the BAP, by introducing two flanking arginines that yield a peptide of an 'MS/MS-friendly' size after trypsin digestion of the BAP sequence. We also included a 7XHis tag in the BAP-domain, to have the option of purifying or/and monitoring the amounts of the target protein regardless of its biotinylation status (Figure 1B, bottom).

To test if the newly designed BAP could be biotinylated by BirA, we cotransfected 293T cells with BirA-GFP and BAP-GFP. Regardless of the fact that GFP can form dimers at high concentrations [20], the main aim of this experiment was not to detect the GFP-GFP interaction, but rather to test the BAP biotinylation, as even background biotinylation due to random collisions between non-interacting proteins could be sufficient to give detectable signal. As seen in Figure 1C, the biotinylation was indeed observed; moreover, as expected, the level of biotinylated BAP-GFP linearly increased with the time of incubation of the cells with biotin. Notably, in parallel contrasfections, the GFP carrying a different BAP sequence (MAGLNDIFEAQKIEWHE), used previously [12], was biotinylated with significantly faster kinetics (the time of half-saturation was 15 min, contrasting with 680 min for the newly designed BAP). Because saturation in the BAP biotinylation levels obscures differences in biotinylation efficiency, the slower kinetics of biotinylation of the new BAP was welcome – given our ultimate goal to correlate BAP biotinylation with protein-protein proximity.

**2. Biotinylation levels are interaction and/or proximity dependent.**

To confirm that our method can detect specific protein-protein proximities, we used several experimental systems described below.

*a. Protein oligomerization (TAP54α vs HP1γ)*

TAP54α (RuvB-like 1) was shown to exist as oligomers, composed of several copies together with the closely related TAP54β (RuvB-like 2) [21, 22]. The heterochromatin proteins HP1 (α, β, γ) are also known to oligomerize through their chromoshadow domains [23, 24], and we typically detected various HP1 forms in our affinity pull-downs using epitope-tagged HP1 proteins (VO, in preparation). On the other hand, no interactions between the HP1 and the RuvB-like proteins had been reported, thus the heterologous cotrasfection with these two proteins (e.g., BirA-HP1γ and BAP-TAP54α) is expected to give close to background levels of biotinylation. Figure 2A (top) shows the biotinylation status of the BAP-TAP54α and BAP-HP1γ, when these proteins were coexpressed with either the BirA-TAP54α or BirA-HP1γ fusions. Whereas the total levels of the BAP fusions are not affected, as judged by Western blotting with the anti-6XHis antibody (Figure 2A, top left), the BAP-TAP54α is appreciably biotinylated when coexpressed with the homologous construct BirA-TAP54α. On the other hand, BAP-HP1γ is significantly more strongly biotinylated when coexpressed with the BirA-HP1γ, as compared to the control cotransfection with heterologous BirA-TAP54α (Figure 2A, top right). In analogous experiments, the related proteins TAP54β and HP1α behaved similarly to the TAP54α and HP1γ, correspondingly (not shown).

In the previous experiment, we could not measure the levels of the BirA-TAP54α and BirA-HP1γ proteins. To rule out the remaining possibility that the observed differences in biotinylation are due to variations in the amounts of these proteins between different transfection samples, both BAP-TAP54α and BAP-HP1γ fusions were now coexpressed in the same cells, along with either the BirA-TAP54α or BirA-HP1γ fusions (Figure 2A, bottom). Despite the presence of comparable amounts of the BAP-TAP54α and BAP-HP1γ fusions in each sample (Figure 2A, bottom left), each protein was more strongly biotinylated in the sample with the homologous BirA fusion (Figure 2A, bottom right).

*b. Binary protein-protein interaction (KAP1 and HP1)*

HP1 strongly interacts with the KAP1 protein, and the domain of KAP1 responsible for the interaction, including a crucial residue, has been identified [25, 26]. We chose the HP1-binding domain of KAP1 (called here KAP1BDwt) together with the point mutant abrogating the interaction (KAP1BDmut) to test if our system can detect specific binary protein-protein interactions (Figure 2B). First, we coexpressed either the wild type or the mutant version of BirA-KAP1BD with BAP-HP1γ. BAP-GFP was used as negative control (Figure 2B, top). Strong biotinylation of BAP-HP1γ was observed only when it was coexpressed with BirA-KAP1BDwt (lane 5). The BirA-KAP1Bdmut construct yielded levels of biotinylation (lane 6) comparable to that of BAP-GFP (lanes 7, 8), most likely reflecting the background levels of labeling.

We also performed a reciprocal experiment, by expressing BAP-KAPBD1wt or BAP-KAP1BDmut, together with the BirA-HP1γ fusion (Figure 2B, bottom). Judging by the anti-6XHis Western, the expression levels of the mutant and wild type fragments of KAP1 were identical (Figure 2B, bottom, top). However, the BAP-KAPBD1wt was biotinylated to a significantly greater extent than the mutant protein (Figure 2, bottom, lane 3 versus 4). Notably, the coexpression of the competitor KAPBDwt (which contains no BAP or His-tag) suppressed this biotinylation (lane 5), whereas the mutant competitor KAPBDmut did not have an effect (lane 7).

c. *Colocalization in the nucleus ('PCNA + H3.1' vs 'PCNA + CenpA')*

Two proteins might not stably interact with each other, yet be colocalized in the cell, which could also increase the biotinylation efficiency. Replication processivity factor PCNA binds directly to p150, the largest subunit of CAF-1, and the two proteins colocalize at sites of DNA replication in cells [27]. CAF-1 is a histone chaperone, which assists in chromatin assembly on replicating DNA, binding directly to the H4/H3 histone dimer [28]. No direct interaction between PCNA and histones has been reported. On the other hand, a specialized H3 histone variant CenpA is deposited on centromeres, in DNA replication independent manner, and thus it is not expected to be present at replication sites [29]. Therefore, despite the absence of direct interaction, one would expect to observe higher proximity between PCNA and canonical H3.1 histone, as compared to PCNA and CenpA. Consistent with these expectations, BAP-H3.1 was biotinylated more efficiently by BirA-PCNA fusion, as compared to BAP-CenpA (Figure 2C). Using fluorescent microscopy, we have also observed that in some cells transfected with 'BAP-H3 + BirA-PCNA', BAP.H3.1 is biotinylated in pattern of bright foci, closely resembling the pattern of replication foci. No such pattern was observed in the 'BAP-CenpA + BirA-PCNA' transfected cells, which is consistent with the notion that increased biotinylation of BAP-H3.1 by BirA-PCNA is due to their colocalization at replication sites in nuclei.

*d. Different subnuclear domains (macroH2A vs H2ABBD)*

As another model example of protein proximity due to similar localization in the cell, we used replacement variant histones. The replacement variant histone H2ABBD has been shown to mark active chromatin (i.e., to be excluded from heterochromatin, such as Barr bodies) [30], whereas the variant macroH2A has been shown to associate with repressed chromatin [31, 32]. One should expect that, on average, different molecules of the histone of the same variant type would be closer to each other in the nucleus as compared to the molecules of different histone type. The BirA- and BAP- fusions of these two proteins were generated and cotransfected in different combinations. Both total nuclear lysate and the chromatin fraction were analyzed by Western blot. In both cases, the coexpression with a homologous BirA fusion significantly increased the efficiency of biotinylation of the BAP-histone fusion (Figure S1, left, and for quantification, Figure S1, right), further supporting the notion that, regardless of physical interaction, similar localization could also be detected with proximity-mediated biotinylation.

*e. Intracellular compartmentalization (GFP and HP1)*

Biotinylation efficiency could also depend on compartmentalization, as sharing common compartment will increase the effective concentration of the BirA fusion near the BAP-target and promote the reaction, even when driven by random collisions between non-interacting proteins. Accordingly, we compared biotinylation of BAP-GFP and BAP-HP1γ when coexpressed with BirA-GFP or BirA-TAP54α fusions. Although TAP54α and HP1γ have not been reported to interact, they both are nuclear proteins, whereas GFP is found in both nucleus and cytoplasm (Fig S2, C). Thus, we expected that BAP-GFP will be less efficiently biotinylated by the BirA-TAP54α, as compared to BAP-HP1γ as a target. Indeed, despite the comparable total levels of BAP-GFP and BAP-HP1γ in the cell, (as α-6XHis Western shows, Figure S2A, top, lane 2), BAP-HP1γ was consistently biotinylated with higher efficiency than BAP-GFP, when these two proteins were coexpressed together with BirA-TAP54α (Figure S2A, bottom, lane 2). In control experiment with BirA-GFP fusion, BAP-GFP was biotinylated with higher efficiency than BAP-HP1γ (Figure S2A, bottom, lane 1), consistent with the notion that the fraction of BirA-GFP excluded from the nucleus was not available for the BAP-HP1γ biotinylation, but could biotinylate the respective BAP-GFP fraction.

**3. MS/MS detection and multiplexing.**

We next tested whether the biotinylated and non-biotinylated forms of the BAP peptide could be detected by mass spectrometry instead of Western blot. We used propionylation to protect the non-biotinylated BAP peptide from tryptic cleavage on the target lysine. This method has been widely used elsewhere – for example, in the analysis of histone modifications [33, 34]. Such an approach allows one to obtain modified and non-modified peptides of comparable sizes, facilitating the interpretation of results.

As seen on the Figure 3A, we can detect both the biotinylated and the propionylated (i.e., non-biotinylated) versions of BAP by LC-MS/MS, as verified from the comparison between the detected mass spectra and the theoretically predicted spectra. Thus, as expected, in addition to the Western analysis, LC-MS/MS can also be used to monitor the biotinylation of the target proteins.

In the context of our approach, an important advantage of mass spectrometry is the ability to distinguish between versions of the same peptide that differ in amino acid sequence. One can use this feature by slightly modifying the sequence of the BAP peptide, in order to design a 'multiplexing' approach that would allow one to test and compare the biotinylation levels of different proteins in the same experiment. To test this possibility, we generated two new versions of the BAP peptide, replacing Val with Phe or Tyr, which have different masses (Figure 3B). The ORF of H2AZ was cloned into these vectors, and the resulting plasmids were cotransfected with BirA-H2A in 293T cells. Afterwards, the transfected cells were mixed, and the 6XHis-tagged histones were purified from the guanidine HCl lysate of the cells by immobilization on Ni-NTA beads. The proteins were digested by trypsin directly on the beads, after which the MRM analysis was used to detect, in the same sample, the biotinylated and propionylated forms of all three peptides (Figure 3C).

An aliquot of the Ni-NTA beads was used to elute the three BAP-H2AZ fusions. As can be seen from the comparison between the LC-MS/MS (Figure 3F) and Western (Figure 3D) analyses, whereas Western blotting fails to distinguish between the different biotinylated proteins because they contribute to the same signal on the autoradiograph, all biotinylated and non-biotinylated forms of the corresponding BAP peptides can be detected separately and distinguished from each other by mass spectrometry (See supplementary figures S3 and S4 for the MS/MS spectra of the corresponding ions). Thus, it should be possible to use this approach to simultaneously detect and compare the interactions of different proteins (e.g., different point mutation variants of the same protein), carrying different versions of BAP, with a single BirA fused protein in the same cells.

Alterations in amino acid sequence can change affinity of a BAP to BirA and thus can affect their biotinylation efficiency regardless of protein-protein proximity. Accordingly, we compared biotinylation levels of different BAPs in the context of the same protein pair "BAP-H2AZ + BirA-GFP". As one can see from the Figure S5, the biotinylation efficiencies do vary for different BAP peptides. The resulting problems with interpretation could be controlled by swapping different BAPs between the tested proteins in reciprocal experiments.

## 4. Stable isotopes.

An alternative approach to multiplexing, free of the problem of variation in biotinylation efficiency between peptides with different sequence, is the use of stable isotopes, a strategy common in quantitative proteomics [35, 36]. To test if the stable isotope labeling scheme is also applicable to our system, we repeated the experiment with TAP54α and HP1γ (Figure 4A). This time, however, the BirA-TAP54α–containing combination was grown in medium containing $^{12}C_6$ lysine, whereas the BirA-HP1γ was grown in medium containing $^{13}C_6$ lysine. After labeling with biotin, the cells were harvested and the two suspensions combined. All further procedures, including SDS-PAGE separation, gel band excision and trypsin digestion were performed with the mixture. The degree of BAP-HP1γ biotinylation by BirA-HP1γ and BirA-TAP54α was estimated by calculating the ratios between the biotinylated versus propionylated versions of the heavy (BirA-HP1γ) or light (BirA-TAP54α) peptides (See Figure 4B for the MS/MS data). As seen on the Figure 4C, a significant difference (about 17 fold) in biotinylation of BAP-HP1γ by the two BirA fusions was observed; the reciprocal difference was seen for the BAP-TAP54α (not shown).

The quantitative measurement of biotinylation should benefit from determining the molar amounts of biotin label on the BAP peptide, which is challenging to estimate with Western analysis, as the signal intensity can depend dramatically on the antibody used. A comparison between the TIC signal of the biotinylated and propionylated BAP in LC-MS/MS also cannot be used directly, as the ionization efficiency generally depends on chemical structure and will thus be affected by the change from a propionyl to a biotin residue. The use of stable isotopes helped us to estimate relative ionization intensities between these two forms of BAP. Cells grown on $^{12}C_6$ lysine were cotransfected with BAP-HP1γ and BirA-HP1γ and then labeled overnight with biotin. In parallel, cells were grown on $^{13}C_6$ lysine medium and transfected with the same plasmids, but not labeled with biotin. The labeled cells were harvested and mixed, in equal amounts, with the corresponding sample of the heavy-isotope–grown cells (Figure 4D). The BAP-HP1γ protein, prepared from the cell mixture by $Ni^{++}$-NTA pulldown and SDS-PAGE separation, was propionylated, digested by trypsin, and analyzed by LC-MS/MS (Figure 4E). The relative ionization coefficient k between the biotinylated and propionylated BAP peptides was estimated as k = (HP - LP)/(LB - HB), where HP corresponds to the area of TIC of heavy propionylated, LP of light propionylated, LB of light biotinylated, and HB of heavy biotinylated BAP. K were calculated for two independent experiments, giving an average value 11.9 ± 1,6 (Figure 4F).

**5. Monitoring the protein at a defined time after the interaction has occurred.**

The distinctive feature of our method is in leaving one of the tested proteins (BAP-fusion) with a permanent molecular mark, which can persist after the proximity between the two proteins has been lost. One can take advantage of this property to add kinetic dimension to the study of protein-protein proximity.

As one such possibility, we tested if a pulse-chase labeling with biotin can be used to trace the fate of a BAP-fused protein of interest that was interacting at the moment of labeling with its BirA-fused proximity partner. This approach is particularly interesting if one wishes to study multi-step intracellular processes, where the interaction partners and protein properties (such as its localization and post-translational modifications) change. Thus, in addition to purifying and studying the properties of a protein of interest at the moment of its interaction/proximity with another protein, our method will enable one also to study the same cellular fraction of the protein at defined times after the interaction has occurred.

As a 'proof of principle' model to explore this opportunity, we used the Rad18 protein, known to form characteristic foci in cell nuclei (these foci are believed to correspond to specialized compartments containing replication proteins, including PCNA). MRC fibroblasts were cotransfected with the BirA-Rad18 fusion together with BAP-H2A, a histone that is homogenously distributed over the nucleus, but, in our system, is expected to be biotinylated only in the neighborhood of the BirA-Rad18 containing foci. Two days after transfection, the cells were pulse-labeled with biotin (Figure 5A). The cells were either fixed immediately after the biotin pulse (pulse sample), or else intensively washed to eliminate biotin and then left in fresh medium for 2 hours before fixation (chase sample). As demonstrated by Western analysis (Figure 5B), only the BAP-H2A histone is biotinylated under these conditions, hence immunofluorescent microscopy detection of the biotin signal should reveal a localization of the BirA-Rad18 in proximity to BAP-H2A at the moment of pulse-labeling.

Consistent with such expectation, staining of the pulse-labeled cells with streptavidin-Cy3 revealed foci of labeled BAP-H2A which colocalized with Rad18 (Figure 5C top) as well as with PCNA (Supplementary Figure S6), detected by specific antibodies. However, the analysis of the chase sample revealed a different pattern of staining (Figure 5C middle, and Supplementary Figure S6). Whereas both Rad18 and PCNA formed foci, the biotin signal appeared more diffuse, and we generally observed a smaller number of foci per cell, whereas the colocalization of the remaining biotinylated foci with the PCNA or Rad18 foci was considerably less pronounced (see Figure 5C bottom for an example).

We conclude that, as was expected, BAP-H2A was biotinylated in the proximity of BirA-Rad18 at the moment of the biotin pulse labeling, and that, with time, the Rad18 partially changed its location inside the nucleus and started to associate with different parts of chromatin, as could be seen by the loss of colocalization of the Rad18 (and PCNA) foci with the biotinylated histones. Regardless of its precise interpretation, this observation clearly demonstrates the possibility to monitor the protein of interest (in this case, BAP-H2A), which was proximity-biotinylated by the BirA fusion of another protein, at later times - i.e., after the proximity between these two proteins could have been lost.

## Discussion

FRET/BRET, PCA and two-hybrid methods – the predominant approaches for studying protein-protein interactions *in vivo* – frequently lie outside the sphere of expertise of a typical proteomics laboratory. The development of more mass-spectrometry and biochemistry-oriented protocols would be a welcome addition to the application portfolio of a standard mass-spectrometry facility.

Also, the FRET/BRET and PCA data are usually interpreted in terms of protein-protein interactions. However, as far as *in vivo* data are concerned, the interpretation in terms of *proximity* between two proteins appears to be more adequate than that of 'physical interaction'. First, unlike the *in vitro* binding studies, where the environment (ion composition, cofactors, etc) can be controlled at will, the intracellular microenvironment is far more challenging to control (to say the least), and the contribution of local intracellular context in stabilizing an interaction can never be ruled out. Second, two proteins could interact via an intermediate - for example, be near each other in cell due to their binding to two closely located sites on DNA. Finally, the role of dynamic self-organization [1, 2] in bringing molecules together, without their physical interaction, also cannot be ruled out. Thus, in the *in vivo* context, the notion of protein-protein proximity (and, accordingly, 'PPP-networks'), as opposed to protein-protein interactions (PPI-networks), appears to be a more adequate term - as, in addition to the direct physical interactions, it can also cover interactions via intermediates, as well as the aspects of intracellular order arisen via dynamic self-organization. Regardless of physical interactions, any information about the protein proximity *in vivo* has a value in itself, as it could be useful in the ultimate task of reconstructing the spatial order in the cell.

In these terms, one shortcoming of the FRET/BRET, PCA and two hybrid approaches is that they measure short-range proximity only. The methodologies that could probe the protein proximities on larger scales could serve as a complement to these established methods, providing additional information about intracellular order. Given that it is sensitive to intracellular compartmentalization, the scope of proximity utilizing biotinylation approach is apparently not limited to the short-range proximities between proteins. We have to admit, however, that the exact distance range in which the proximity can still be detected could depend on the pair of proteins studied and thus merits a further study. In particular, the distance range can depend on the mutual orientation and the mobility of proximity partners. As an illustration (Figure S7), first consider two proteins ($A_1$ and $A_2$) that are close to each other, but their movement is strongly constrained (e.g., by being tethered to a rigid scaffold). In this case, the BirA and BAP moieties have a low probability to directly encounter each other. In the second case, the average distance between proteins ($B_1$ and $B_2$) is larger, but they move more freely. This will lead to higher chance of biotinylation, which, as a permanent modification, will accumulate on the BAP-tagged protein. This illustration, together with the effects of compartmentalization, also suggests that due to the many factors that can contribute to the differences in observed biotinylation levels, the correct interpretation of data requires careful choice of proper controls. For example, when testing proximity between nuclear proteins, one cannot use BAP-GFP as a negative control, as its extranuclear fraction will be inaccessible for BirA fused to a nuclear protein and thus contribute to an artificially low background biotinylation levels (and any pair of nuclear proteins will give a higher biotinylation signal). The proper negative control should be BAP-GFP with a nuclear localization signal. On the other hand, protein-protein interactions or proximities are often studied not for their own sake, but rather to access the network of causal relations in the cell. From this perspective, the increased ability of two PUB proteins to biotinylate one another (above the rate expected from the modeling of biotinylation rates based on random protein-protein collisions in a homogenous solution) has direct relevance, regardless of the exact interpretation of what had caused the biotinylation increase.

To demonstrate the utility of our approach, we have chosen several examples of protein pairs that represent different instances of protein proximity. a). Many proteins can form homooligomers, and we demonstrated, using the examples of TAP54$\alpha$ and HP1$\gamma$ self-association, that such interactions can be detected with our method. b). To show that this methodology is not limited to protein self-association, we also demonstrated that it can detect interaction between two different proteins (e.g., KAP1 and HP1$\gamma$). c). Two proteins do not have to interact directly, but can be in proximity in common locus (e.g., PCNA and histone H3.1, or the alternative histones H2ABB and macroH2A, enriched in active chromatin and repressed chromatin in the nucleus, respectively). d) Finally, sharing common intracellular compartment will increase the effective concentration of the BirA-fusion near the BAP-target and promote the reaction, as we demonstrated using GFP and HP1 as examples of compartmentalization differences.

Another type of additional information that could be provided by PUB is based on the fact that biotinylation leaves one of the proteins with a permanent modification, which persist after the protein-protein proximity could have been lost. We presented a way to take advantage of this feature that adds kinetic dimension to the studies of the protein-protein proximity. Using BAP-fused histone H2A and BirA-Rad18 as a model, we demonstrated that one can use a pulse-chase labeling with biotin to trace the fate of a BAP-fused protein of interest that was interacting with a given BirA-fused protein at the moment of labeling. Among the many potential applications of such type of analysis are the study of protein transport (e.g., by using a fusion of BirA with a component of the membrane pore complex and labeling a protein transported to the nucleus by BAP), measurement of the stability and consequent modifications of a particular fraction of protein that has been a part of a complex with a specific protein partner, etc. Thus, in addition to helping to reconstruct the cell structure in space, our approach has a potential to add temporal dimension to such reconstruction studies.

We also need to mention a difference of PUB approach from the other biochemistry-friendly approaches to study protein-protein interactions, such as co-immunoprecipitation and TAP (tandem-affinity purification). Our method is intended for *monitoring* interactions/proximities between two given proteins - not for *identification* of new interaction partners of a given protein. One can easily envision many instances when such a task might be needed. A case in point is structure-functional analysis of a protein (deletion and/or site-specific mutagenesis) with the aim to map interaction domains (as we illustrated on the example of the KAP1 wild type and mutant interacting with HP1). This limitation notwithstanding, our method has the following appealing difference from TAP and related approaches to study protein-protein interactions. These methods work satisfactorily only for the sufficiently strong interactions that can withstand the procedures of solubilization and affinity purification. For example, the study of membrane complexes or the interactions between matrix proteins is challenging with these methods, given the severe limitations on the extraction and purification conditions compatible with the stable protein-protein interactions. On the contrary, in the proposed approach, the generation of a permanent covalent mark on one of the proteins studied will allow one to bypass the limitations imposed by the extraction and purification procedures. Thus the method should prove useful for the study of interactions that are otherwise difficult to detect by the Co-IP and TAP methods.

In some cases, e.g., when comparing interactions of two proteins between different cell lines with uncontrolled levels of their endogenous counterparts, a potential problem with interpretation could arise due to the competition of the endogenous proteins for the interaction. As seen in Figure 2B, we could efficiently decrease the biotinylation of BAP-KAP1 by overexpressing untagged KAP1. This problem should be taken into account when concluding from results obtained using this method. Notably, however, this valid concern equally applies to other methods to study protein-protein interactions in homologous systems in vivo, such as FRET, BRET and two hybrid (e.g., mammalian two hybrid used to study mammalian interaction partners).

As far as future development of this system is concerned, the vectors utilized in our work require "classic" cloning to be performed for each pair of proteins tested, which greatly increases the amount of work required per experiment, effectively preventing large-scale study. Making our vectors Gateway compatible will facilitate the molecular biology manipulations and benefit from large ORF collection compatible with this tagging system.

Finally, while this project was in progress, we learnt that a somewhat similar approach to detect protein-protein interactions *in vivo* has been developed by another group [37]. However, we wish to emphasize the crucial and distinguishing feature of our system: its mass-spectrometry-oriented format. In this respect, the sequence of the BAP domain used in [37] is not expected to yield a tryptic peptide that would be efficiently detected by mass-spectrometry, thus precluding its use in techniques such as multiplexing with varying amino acid sequences or stable isotope labeling and quantification, which constitute one of the main advantages of the technique we describe here.

**Acknowledgements**

We thank Dr. M. Lechner (Drexel University, Philadelphia) for the KAP1-HP1BDwt and KAP1-HP1BDwt plasmids, Drs H. Willard and M. Chadwick (Case Western Reserve University, Cleveland) for the ORFs of the H2A.BBD and macroH2A, and Dr. L. L. Pritchard for critical reading of the manuscript. This work was supported by grants from "La Ligue Contre le Cancer" (9ADO1217/1B1-BIOCE), the "Institut National du Cancer" (247343/1B1-BIOCE) and Centre National de la Recherche Scientifique [CNRS-INCA-MSHE Franco-Pologne #3037987) to VO, by NCB Kazakhstan (0103_00404) to AK and by a travel grant to A.K. from the Coopération Universitaire et Scientifique Department of the French Embassy in Astana, Kazakhstan.

**Supporting Information Available:** This material is available free of charge via the Internet at http://pubs.acs.org.

**Figure legends**

**Figure 1. Design and features of the system.**

**A. The principle.** Two candidate interaction proteins X and Y are coexpressed in the cells of interest, the first protein being expressed as a BirA fusion, and the second one as a fusion to the Biotin Acceptor Peptide (BAP). The cells are pulse-labeled with biotin. Where there is interaction (top), the BAP-X fusion should be more efficiently biotinylated.

**B. Vector design**. Top – the positions of the CMV/MoMuLV enhancers, and of the BAP and BirA sequences, relative to the cloned ORF, are indicated. Bottom – nucleotide and amino acid sequences encoding the newly designed His-BAP are shown.

**C. New BAP is biotinylated with slower kinetics as compared to the old BAP**. Left – Western analysis of biotinylation of the new BAP-GFP (ILEAQKIVR, left) versus previously used efficient old BAP-GFP (MAGLNDIFEAQKIEWHE, right) in the presence of BirA-GFP fusion. (top) – α-6XHis-HRP signal, (bottom) – streptavidin-HRP signal. Right – quantification of the kinetics of biotinylation of the two BAPs. The streptavidin signal was first normalized by taking a ratio with the respective α-6XHis signal. For every individual experiment, the normalized value of the biotinylation level for each time point was divided by the respective value for the 24hr time point from the same experiment, which was taken as 100%. Plotted are averaged results from three independent experiments.

**Figure 2. Biotinylation levels are interaction/proximity dependent.**

**A. HP1γ versus TAP54α oligomerization.**

**Top.** 4 combinations of BirA and BAP fusions were transfected separately into cells: 1,5 – BirA-TAP54α + BAP-TAP54α; 2,6 - BirA-HP1γ + BAP-TAP54α; 3,7 - BirA-TAP54α + BAP-HP1γ; 4,8 - BirA-HP1γ + BAP-HP1γ. 1-4 – α-His-HRP Western, 5-8 – streptavidin-HRP Western. The positions of the BAP-fusions and nonspecific signal (NS) are indicated by lines, and the biotinylated proteins are indicated by black triangles.

**Bottom.** Biotinylation of BAP-HP1γ and BAP-TAP54α, cotransfected into the same cells, by either BirA-HP1γ or BirA-TAP54α. 1,4 - untransfected cells; 2,5 – BirA-TAP54α + BAP-TAP54α + BAP-HP1γ; 3,6 - BirA-HP1γ + BAP-TAP54α + BAP-HP1γ. Left – α-His-HRP Western, Right – streptavidin-HRP Western. The positions of the BAP-fusions (triangles) and nonspecific signal (NS) are indicated.

Importantly, the nonspecific signal (NS) is also detected in untransfected cells, thus it corresponds to cellular proteins that bind to the α-His antibodies or streptavidin, and do not represent nonspecific biotinylation by BirA-fusions.

**B. Binary interaction between HP1γ and KAP1.**

**Top.** BirA-KAP1BD biotinylating BAP-HP1γ. 1,5 - BirA-KAP1BDwt + BAP-HP1γ ; 2,6 - BirA-KAP1BDmut + BAP-HP1γ; 3,7 - BirA-KAP1BDwt + BAP-GFP; 4,8 - BirA-KAP1Bdmut + BAP-GFP. Left – α-His-HRP Western, Right – streptavidin-HRP Western. The positions of the BAP-fusions and nonspecific signal (NS) are indicated.

**Bottom.** BirA-HP1γ biotinylating BAP-KAP1BD. 1,3,5,7 – BirA-HP1γ + BAP-KAP1BDwt; 2,4,6,8 – BirA-HP1γ + BAP-KAP1Bdmut. 1,2 – no biotin was added to cells; 5,6 – 10-fold excess of a plasmid expressing untagged KAP1BDwt was added, 7,8 – the same as 5,6, but KAP1BDmut used. Top – α-His-HRP Western, Bottom – streptavidin-HRP Western.

**C. Colocalization in the nucleus (PCNA+H3.1 vs PCNA+CenpA)**

**Left.** Western analysis of the biotinylation efficiency of BAP-H3.1 (lanes 1,3) as compared to BAP-CenpA (2,4) by BirA-PCNA fusion (3,4). BirA-GFP biotinylation (1,2) was used to control for the differences in biotinylability of BAP-peptide in context of H3.1 and CenpA fusion. Top – αHis, Bottom – Streptavidin signals.

**Middle**. Quantification of the relative biotinylation efficiency. The signal intensities were first measured by densitometry, then the streptavidin signal for every BAP-fusion was normalized by dividing it to the α-His signal. Relative efficiency of biotinylation of a respective BAP-fusion by the BirA-PCNA as compared to BirA-GFP was obtained by dividing the normalized values of biotinylation for the lanes 3 and 1 for H3.1, and for lanes 4 and 2 for CenpA, respectively. Shown are results of calculations for two independent experiments. The lower biotinylation signal in the case of BirA-PCNA transfection is due to a significantly lower levels of BirA-PCNA expression levels as compared to BirA-GFP (not shown).

**Right.** Replication-like foci of biotinylated BAP-H3.1 after cotransfection with BirA-PCNA. Confocal microscopy of cells cotransfected with "BAP-H3.1 + BirA-PCNA" (top) or "BAP-CenpA + BirA-PCNA" (bottom), labeled with biotin 48 hr after transfection, and stained with straptavidin-Cy3 (left) or analyzed by Differential Interference Contrast (DIC, right).

**Figure 3. Multiplexing.**

A. **MS/MS detection of BAP.** Shown are the MS/MS fragmentation spectra of the propionylated (top) and biotinylated (bottom) forms of BAP, with the detected y-series and b-series ions indicated.

B. **New BAPs.** Shown are the nucleotide and amino acid sequences of the new BAPs (BAP1118 and BAP1135). The site of the amino acid change, as compared to the original BAP1070, is indicated by a box.

C. **Scheme of the multiplexing experiment.** Cells were transfected with BirA.H2A and with a vector expressing different BAP fusions of H2AZ. After 4 h labeling, they were harvested, mixed, and the BAP-H2AZ proteins purified on $Ni^{2+}$-NTA agarose.

D. **$Ni^{2+}$-NTA purification of the BAP-H2AZ fusions.** M- Marker, Input – input, FT – flowthrough, E – eluate. Left – Coomassie Brilliant Blue stain, right – streptavidin-HRP Western. The positions of BAP-H2AZ and the ubiquitinated BAP-H2AZ are indicated.

**E.   MRM detection of different BAPs.** Shown are extracted ions chromatograms for the most intensive fragmentation ions present in the MS/MS spectra of the respective peptides. Note that the BAP1135 is propionylated on both K and Y, which affects the M/Z of the respective ions.

**Figure 4. Stable isotopes.**

**A. Experimental scheme 1.** Cells were grown in medium containing $K^{12}C_6$ or $K^{13}C_6$ and transfected, respectively, with BirA-TAP54α or with BirA-HP1γ, along with a mixture of BAP-TAP54α and BAP-HP1γ. After 4 h labeling, the cells were harvested and mixed, the proteins from the nuclei were fractionated on SDS-PAGE, and the gel bands containing the HP1γ and TAP54α were processed for MRM analysis on LC-MS/MS.

**B. MS/MS spectra of heavy BAP1070.** Shown are the MS/MS fragmentation spectra of the propionylated (top) and biotinylated (bottom) forms of $K^{13}C_6$-labeled BAP1070, with the detected y-series and b-series ions indicated.

**C. MRM analysis of a SILAC experiment.** Shown are extracted ions chromatograms (EIC) for the most intensive fragmentation ions present in the MS/MS spectra of the BAP1070. PL - propionylated light, PH - propionylated heavy, BL - biotinylated light, BH - biotinylated heavy. The areas of every peak (A) are indicated on the left of the corresponding chromatogram. PL and BL correspond to transfection with the BirA-TAP54α fusion, whereas PH and BH correspond to transfection with the BirA-HP1γ fusion; accordingly, the BAP1070 peptide is 17 times better biotinylated in the BirA-HP1γ–fusion-containing sample.

**D. Experimental scheme 2.** Cells were grown in medium with $K^{12}C_6$ or $K^{13}C_6$ and transfected with BirA-HP1γ and BAP-HP1γ. The $K^{12}C_6$ grown cells were labeled overnight with biotin before being harvested and mixed, in equal amounts, with the $K^{13}C_6$ cells grown in absence of biotin.

**E. MRM quantification of different forms of BAP1070.** Shown are extracted ion chromatograms for the different forms of BAP1070 from the above experiment. The positions of the peaks (labeled as in the legend to Fig. 4c) are indicated by arrows. Control - the $K^{12}C_6$ grown cells were not labeled with biotin. Biotin - the $K^{12}C_6$ grown cells were labeled with biotin.

**F. Determination of the coefficient of relative ionization of the propionylated and biotinylated forms of BAP1070**

Shown are the histograms representing the areas of the peaks from the control experiment (as in Fig. 4E) and two independent biotinylation experiments (Biotin 1 and Biotin 2), and the relative ionization coefficient k between the biotinylated and propionylated BAP peptides for each experiment, estimated as k = (HP - LP)/(LB - HB), where the areas of the peaks are labeled as in the legend to Fig. 4c.

**Figure 5. Pulse-chase experiment**.

**A. Experimental scheme.** MRC fibroblasts are cotransfected by the BirA-Rad18 fusion together with BAP-H2A. Two days after transfection, the cells are pulse-labeled with biotin 6 h later. The cells are either fixed immediately after the biotin pulse (pulse sample), or else washed extensively to remove biotin, then left for 2 h before fixation (chase sample).

**B. Specificity of biotinylation.** Western blot analysis with streptavidin-HRP, showing that a specific signal appears only after BirA-Rad18 and BAP-H2A are coexpressed.

1 - control untransfected sample, 2 - BirA-Rad18, 3 - BirA-Rad18 + BAP-H2A, 4 - BAP-H2A. The two forms of H2A (non-ubiquitinated and ubiquitinated), both biotinylated in this experiment, are indicated by asterisks.

**C. Colocalization analysis.** Left Top - pulse sample. Middle - chase sample. Bottom – zoomed area from the chase sample showing an example of biotinylated foci that do not colocalize with the Rad18 foci and vice versa (indicated by asterisk).

**Table of Contents (TOC) Graphic & Synopsis**

We have developed PUB-MS (for Proximity Utilizing Biotinylation and Mass Spectrometry), an approach to monitor protein-protein proximity, based on biotinylation of a protein fused to a biotin-acceptor peptide (BAP) by a biotin-ligase, BirA, fused to its interaction partner. We demonstrate the advantage of mass spectrometry by using BAPs with different sequences in a single experiment (allowing multiplex analysis) and by using stable isotopes.

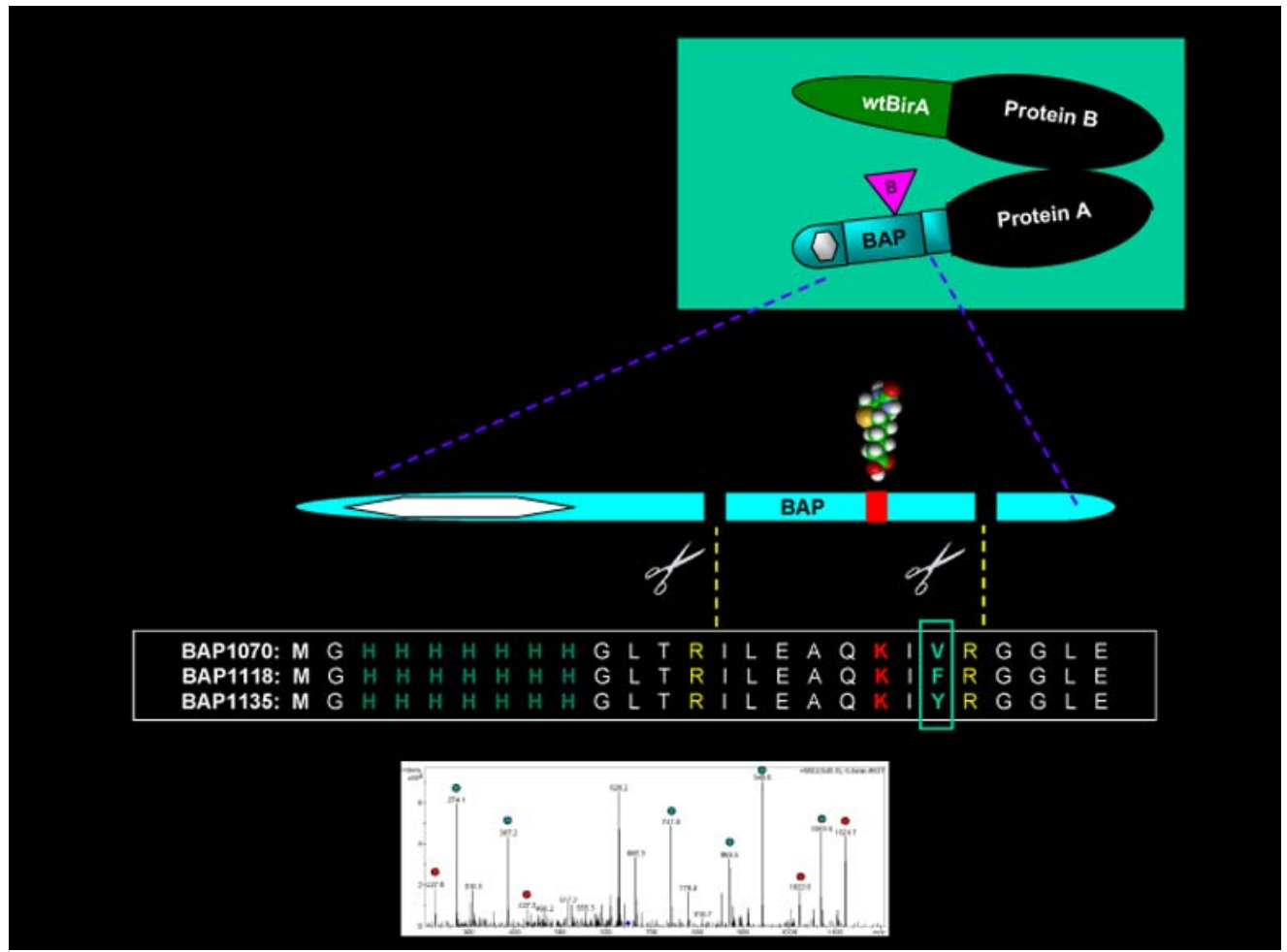

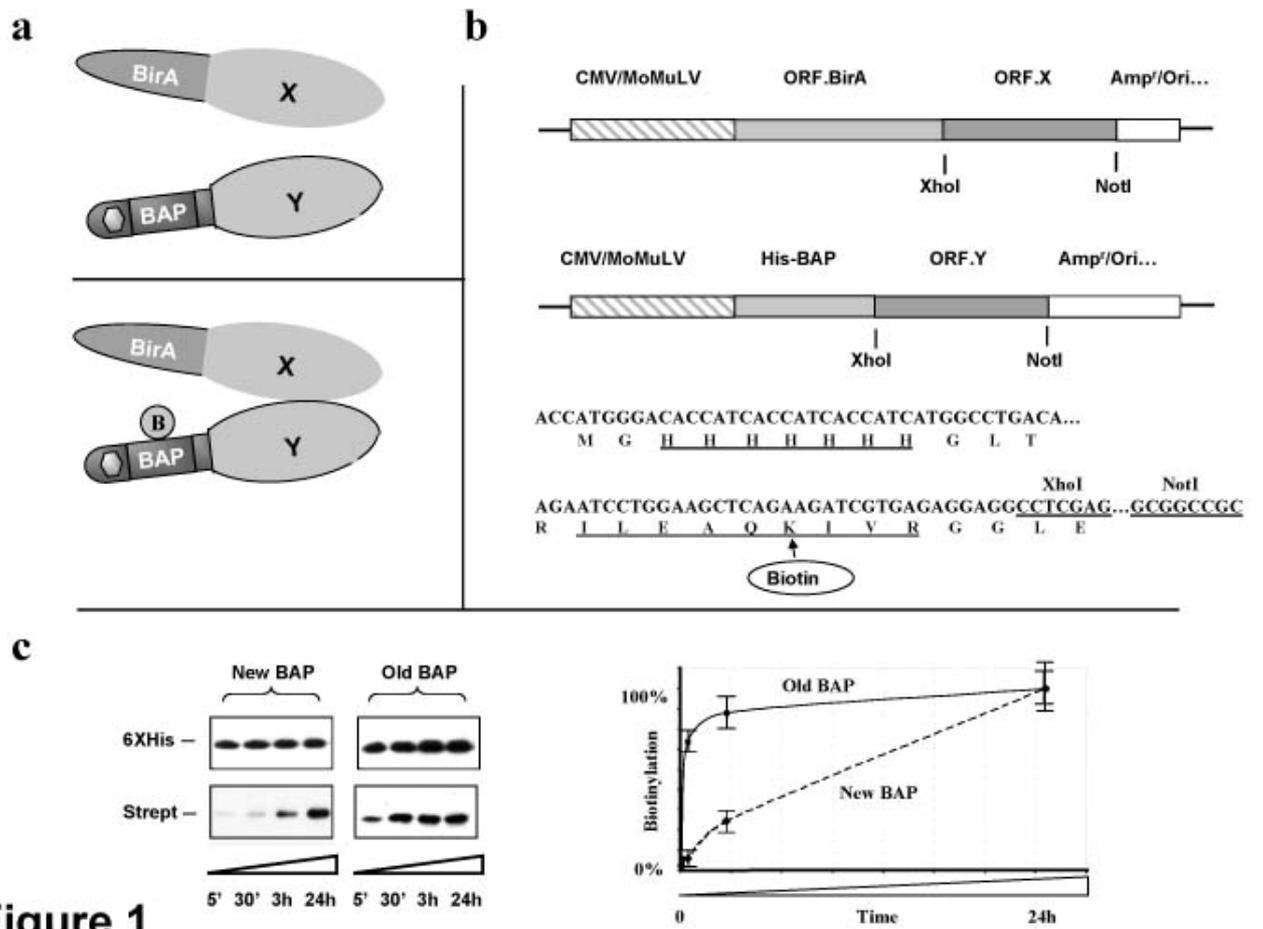

Figure 1
35

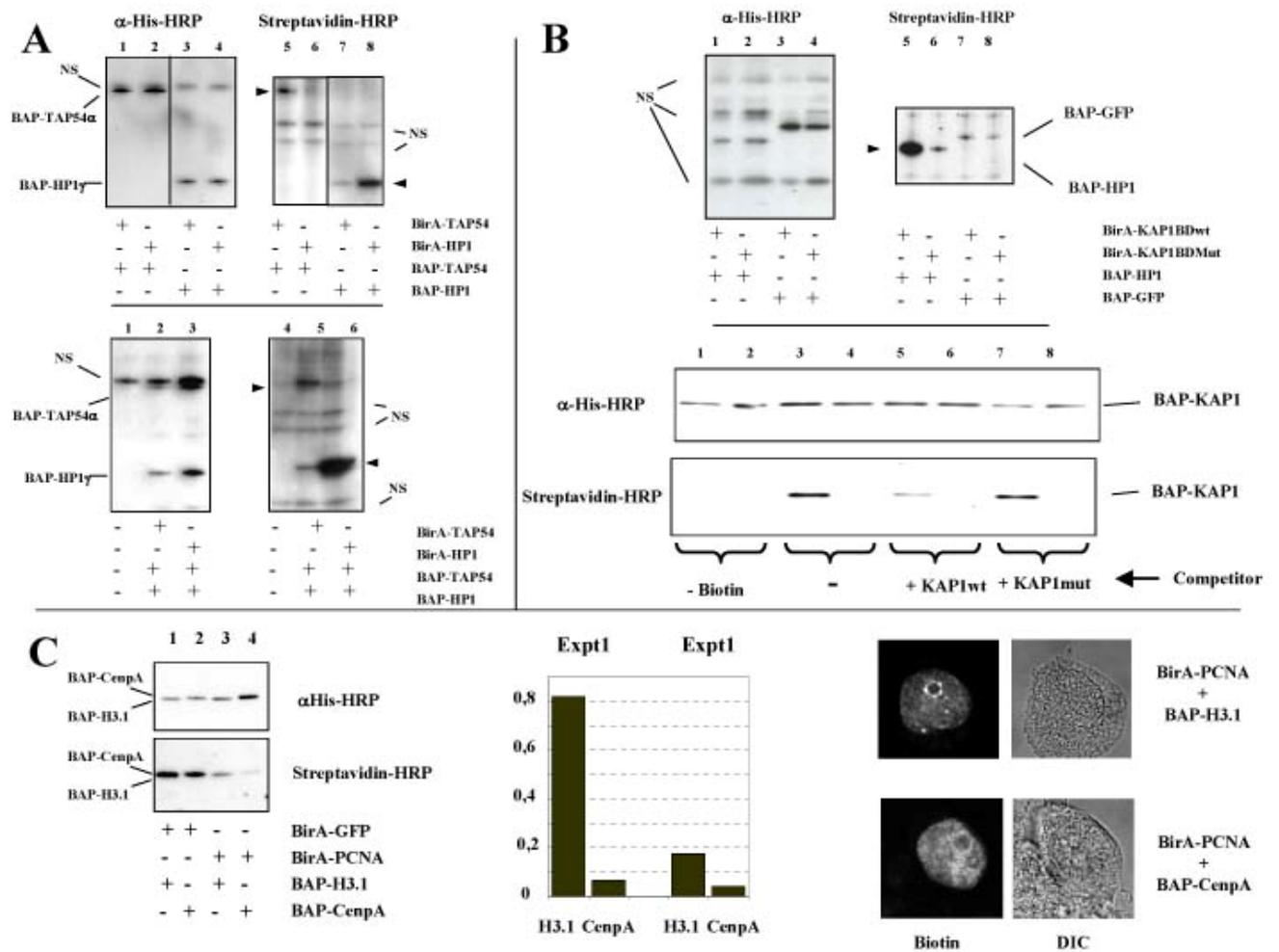

**Figure 2**

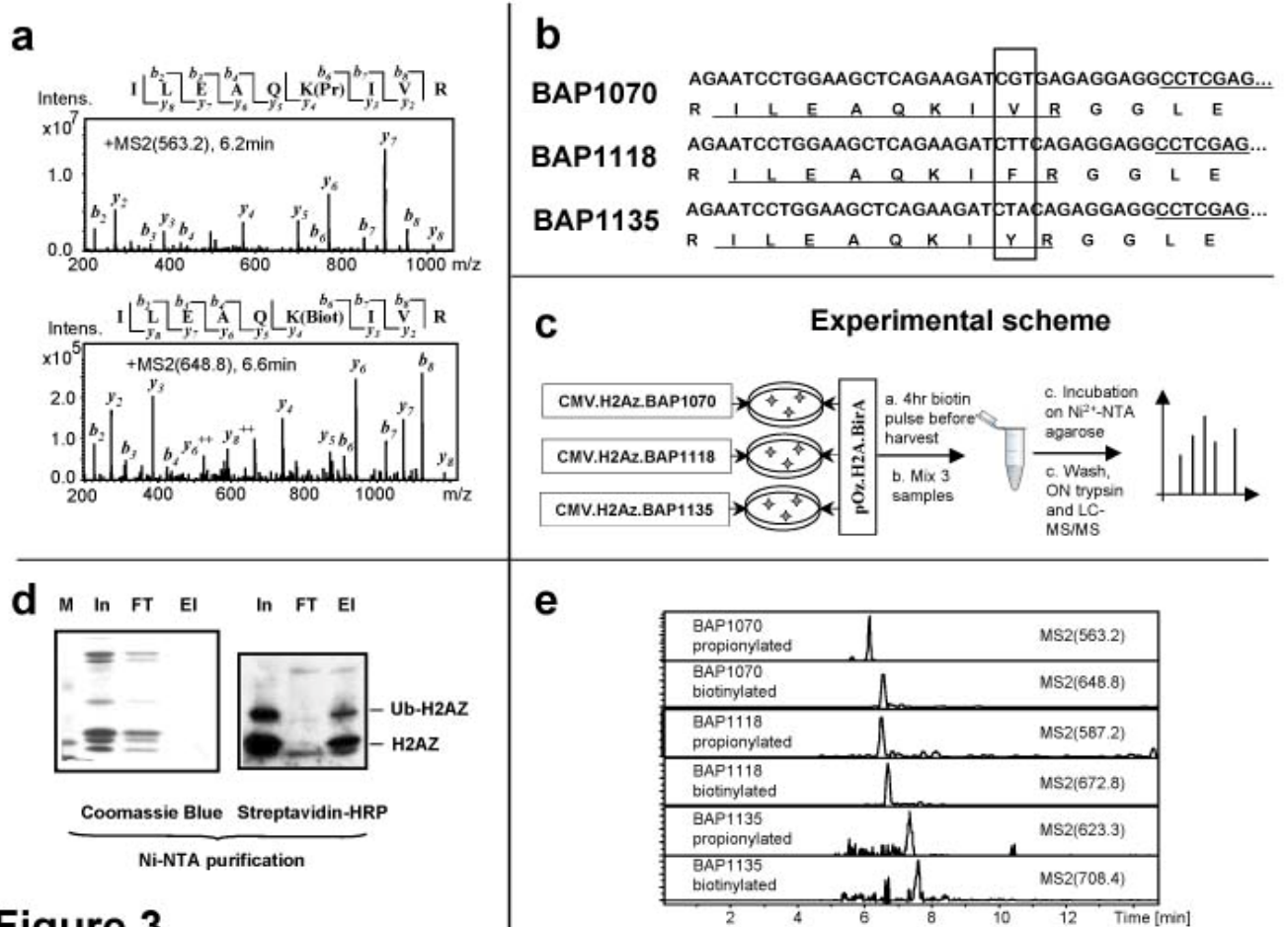

Figure 3

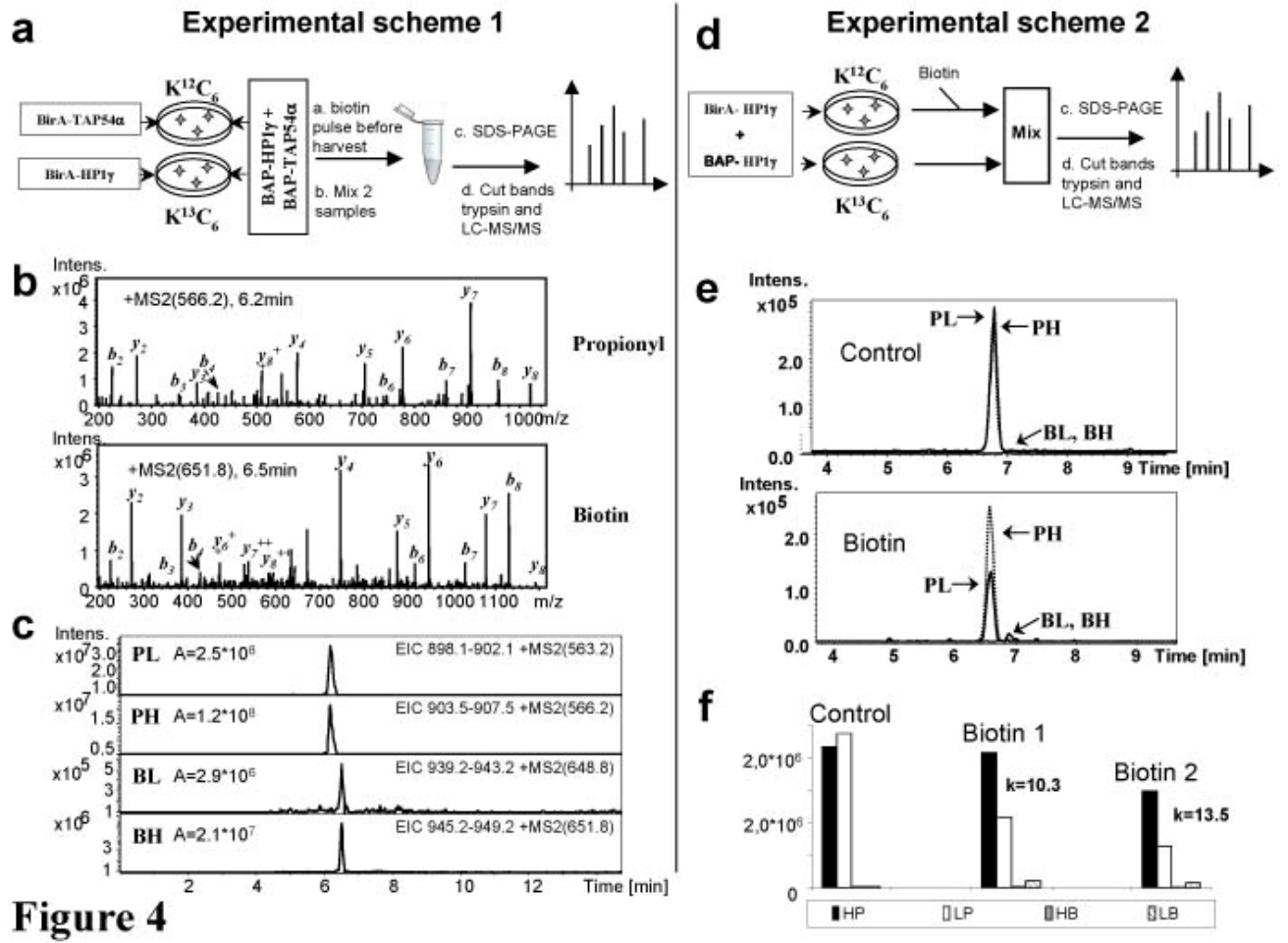

Figure 4

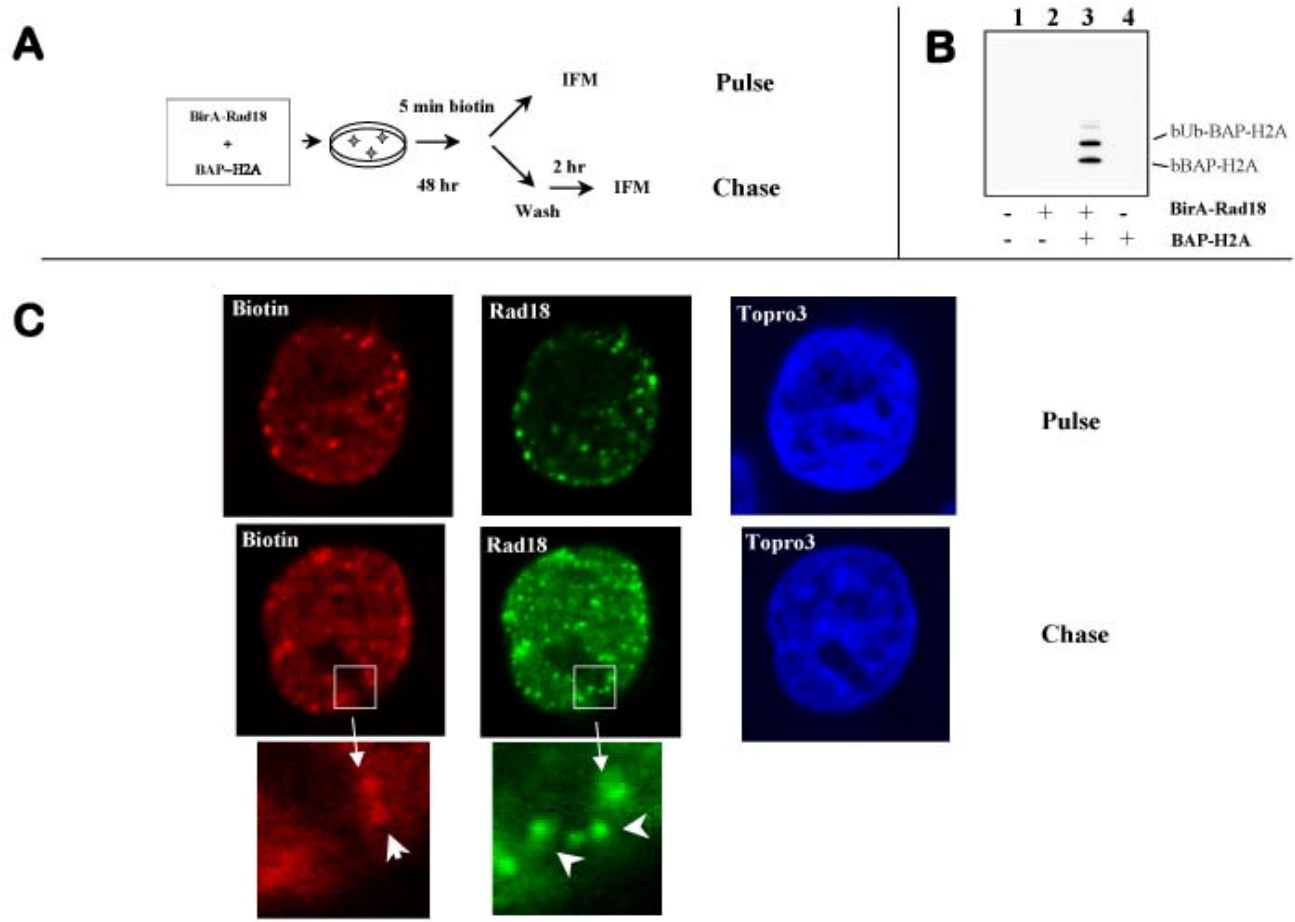

**Figure 5**

**Supplementary materials:**

**Figure S1. macroH2A vs H2ABBD.**

**Left**: 2 combinations of BirA and BAP fusions were transfected separately into cells: 1 – Untransfected cells; 2,4 – BirA-macroH2A + BAP-macroH2A + BAP-H2ABBD; 3,5 – BirA-H2ABBD + BAP-macroH2A + BAP-H2ABBD. 2,3 – total nuclear lysate (Tot), 4,5 – Chromatin associated histones (Chr). Top – α-His-HRP Western, Bottom – streptavidin-HRP Western. The positions of the BAP-fusions and nonspecific signal (NS) are indicated.
**Right**: Quantification of the biotinylation efficiencies. The signal intensities were first measured by densitometry, then the streptavidin signal for every BAP-fusion was normalized by dividing it to the α-His signal. G – the ratio between biotinylation of BAP-macroH2A and BAP-H2ABBD in the presence of BirA-H2ABBD; H – the ratio between biotinylation of BAP-macroH2A and BAP-H2ABBD in the presence of BirA-macroH2A. 1 – total nuclei, 2 – chromatin fraction. **Far right** – Ratio between the efficiencies of heterologous (H2ABBD vs macroH2A) and homologous (H2ABBD vs H2ABBD and macroH2A vs macroH2A) biotinylation. Shown are average values of G/H calculated for three experiments, for total nuclei (1) and chromatin fraction (2)

**Figure S2. Detection of differences in compartmentalization**

**A.** BAP-GFP and BAP-HP1γ where coexpressed either with BirA-GFP (left) or BirA-TAP54α (right) fusions. Top – α-His-HRP Western, Bottom – streptavidin-HRP Western. The positions of the BAP-fusions are indicated.

**B**. Quantification of biotinylation differences. The value of the streptavidin-HRP signal was normalized by taking a ratio with the value of α-His-HRP signal, which reflects the amount of the transfected protein regardless of its biotinylation status. Left: The ratios between the normalized signals of biotinylated BAP-HP1 and BAP-GFP was calculated for both BirA-GFP (G) and BirA-TAP54α (T) cotransfections. Right – the average ratio and standard deviation between the G and T values from 3 independent experiments.

**C.** Intracellular localization of the ectopically expressed HP1, GFP and TAP54. After cotransfection with BAP-GFP, BAP-HP1γ and BirA-TAP54α, the 293T cells were disrupted in CSK buffer, and nuclei pellets (right) were separated from supernatant (left). The volumes of both fractions corresponding to equal numbers of cells were loaded on SDS-PAGE gel, the proteins separated and the presence of BAP- and BirA fusions detected by α-His-HRP Western. Due to the lower expression levels of TAP54 protein (weak retroviral MoMuLV enhancer), significantly longer times of exposure were required to detect the respective signal.

**Figure S3. MS/MS spectra of the biotinylated and propionylated BAP1118** (V→F)**.**

Shown are the MS/MS fragmentation spectra of the propionylated (top) and biotinylated (bottom) forms of BAP, with the detected y-series and b-series ions indicated.

**Figure S4. MS/MS spectra of the biotinylated and propionylated BAP1135** (V→Y).
Shown are the MS/MS fragmentation spectra of the propionylated (top) and biotinylated (bottom) forms of BAP, with the detected y-series and b-series ions indicated. Note that, in this peptide, both Lysine (K) and thyrosine (Y) are propionylated, which affects the masses of the parental and daughter ions.

**Figure S5. Relative biotinylation efficiencies of 3 BAP peptides.**

Western analysis of biotinylation levels of three BAP-H2AZ fusions, cotransfected separately with BirA-GFP. Shown is one example of the analysis. Top – α–His signal, used for normalization. Bottom – Streptavidin signal. The intensities of the signals were measured by densitometry (left tables, top and bottom). First, the values of streptavidin signals were normalized by taking into account different α–His signal intensities (right table). Afterwards, the relative intensities were calculated by dividing the normalized values for BAP1118 or BAP1135 (1,3) by the normalized value for BAP1070 (2). The average values from three different experiments and standard deviations are shown on the right side.

**Figure S6. Colocalization analysis of PCNA and Biotin label after pulse and 2 hr chase.**

Top - pulse sample. Bottom - chase sample.

**Figure S7. Partner mobility can contribute to the biotinylation efficiency**

**Left** - The average positions of the BAP and BirA fusions ($A_1$ and $A_2$, respectively) are close to each other, but their motion is so constrained that they practically have no chance to encounter each other (as depicted by the absence of an overlap between the distributions of their positions).

**Right** - The average distance between two proteins ($B_1$ and $B_2$) is larger than that for the A pair, but their positions fluctuate more significantly, leading to the larger overlap in their positions and thus to accumulation of biotinylated BAD species.

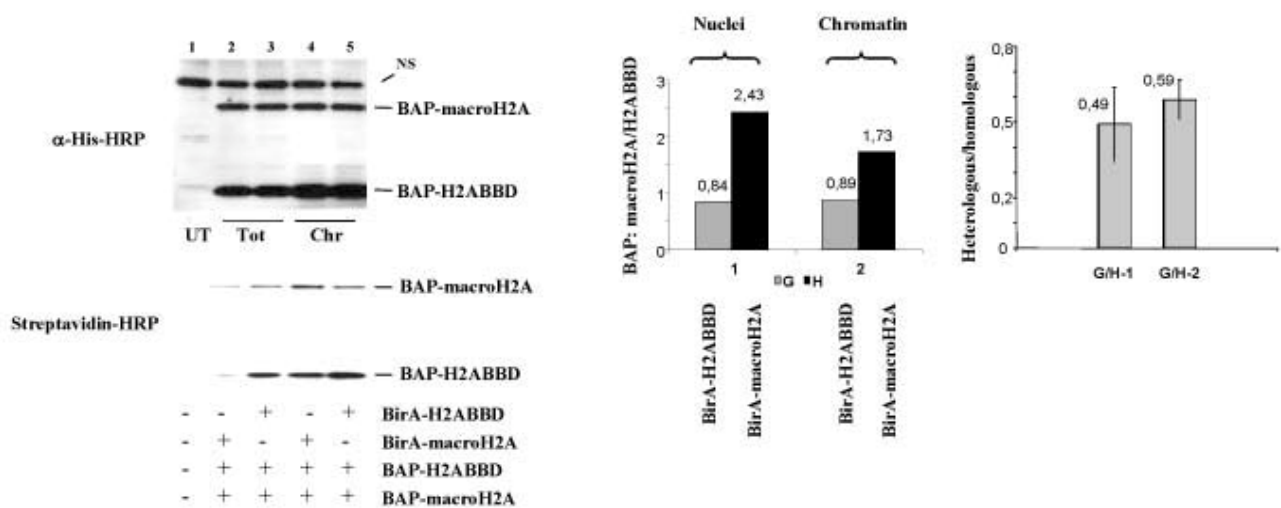

**Figure S1**

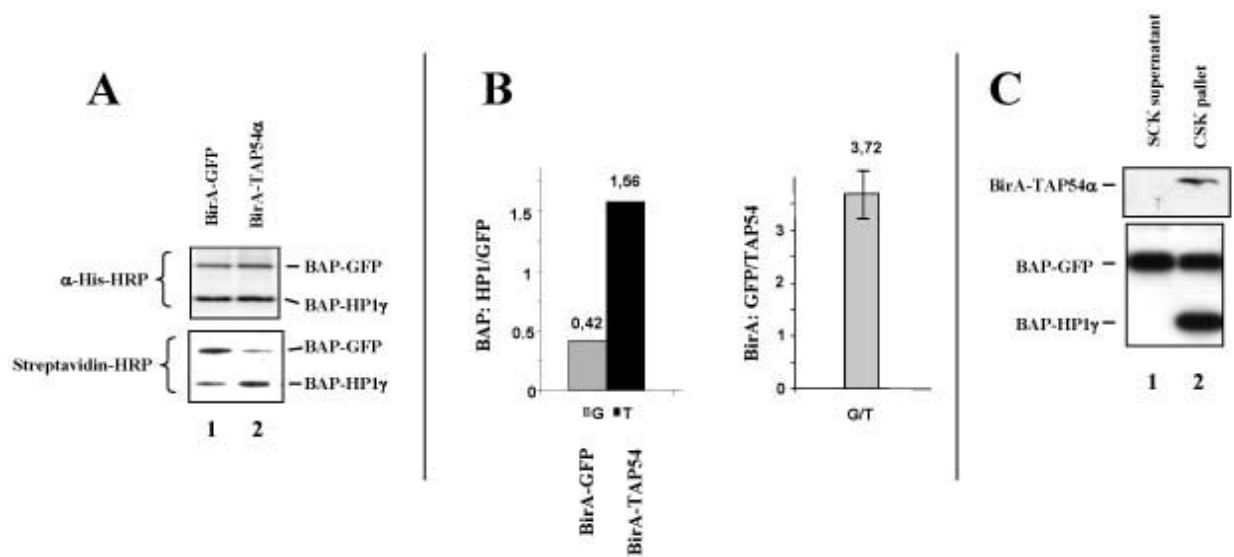

**Figure S2**

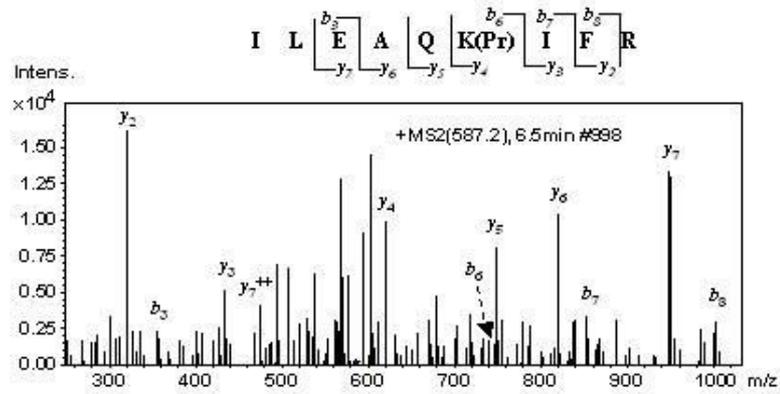
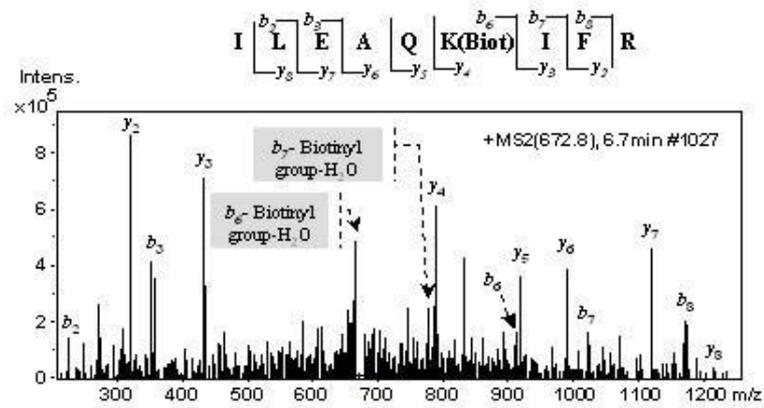

**Figure S3**

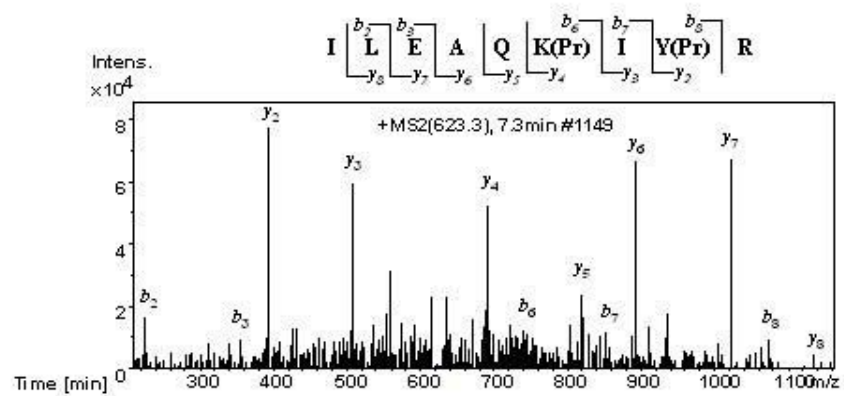
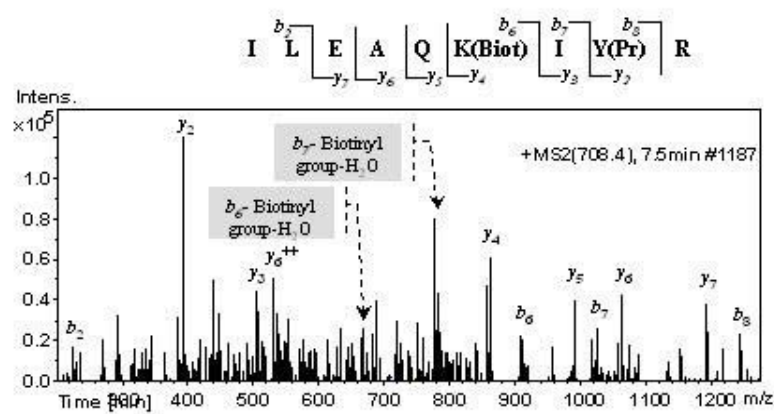

**Figure S4**

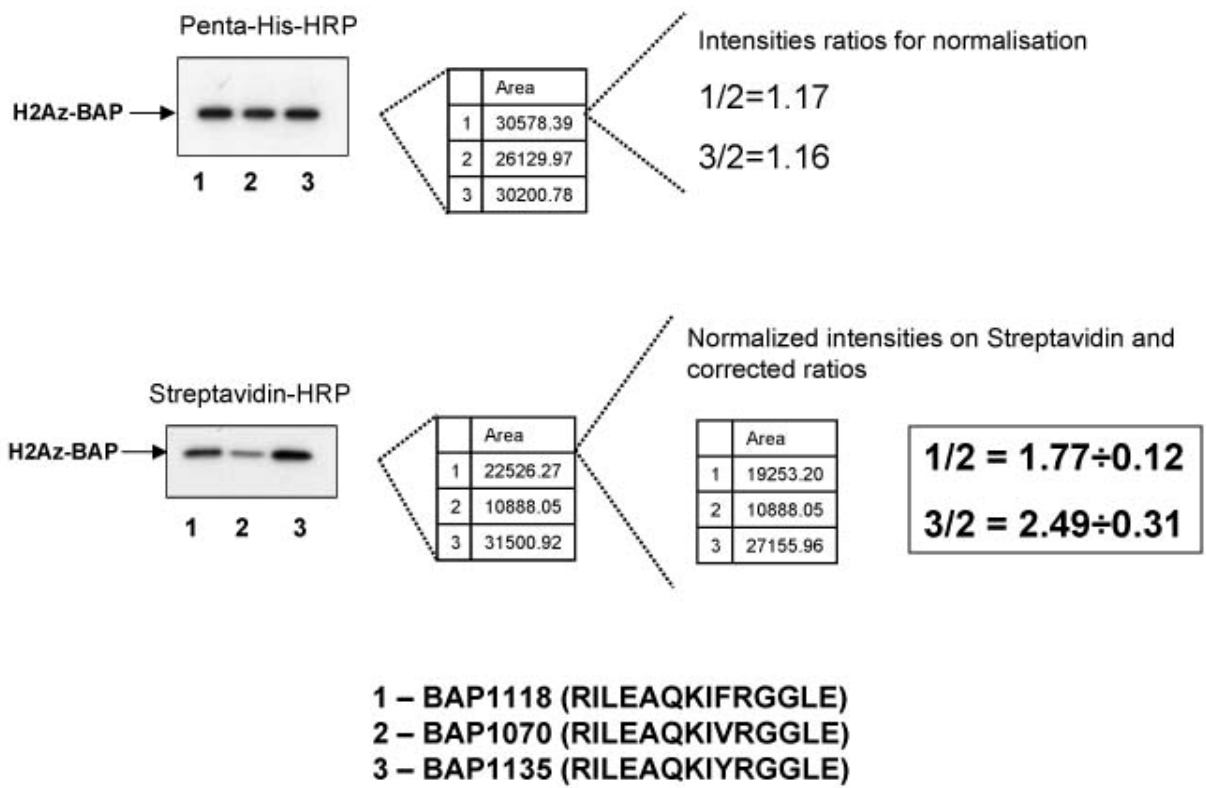

1 – BAP1118 (RILEAQKIFRGGLE)
2 – BAP1070 (RILEAQKIVRGGLE)
3 – BAP1135 (RILEAQKIYRGGLE)

**Figure S5**

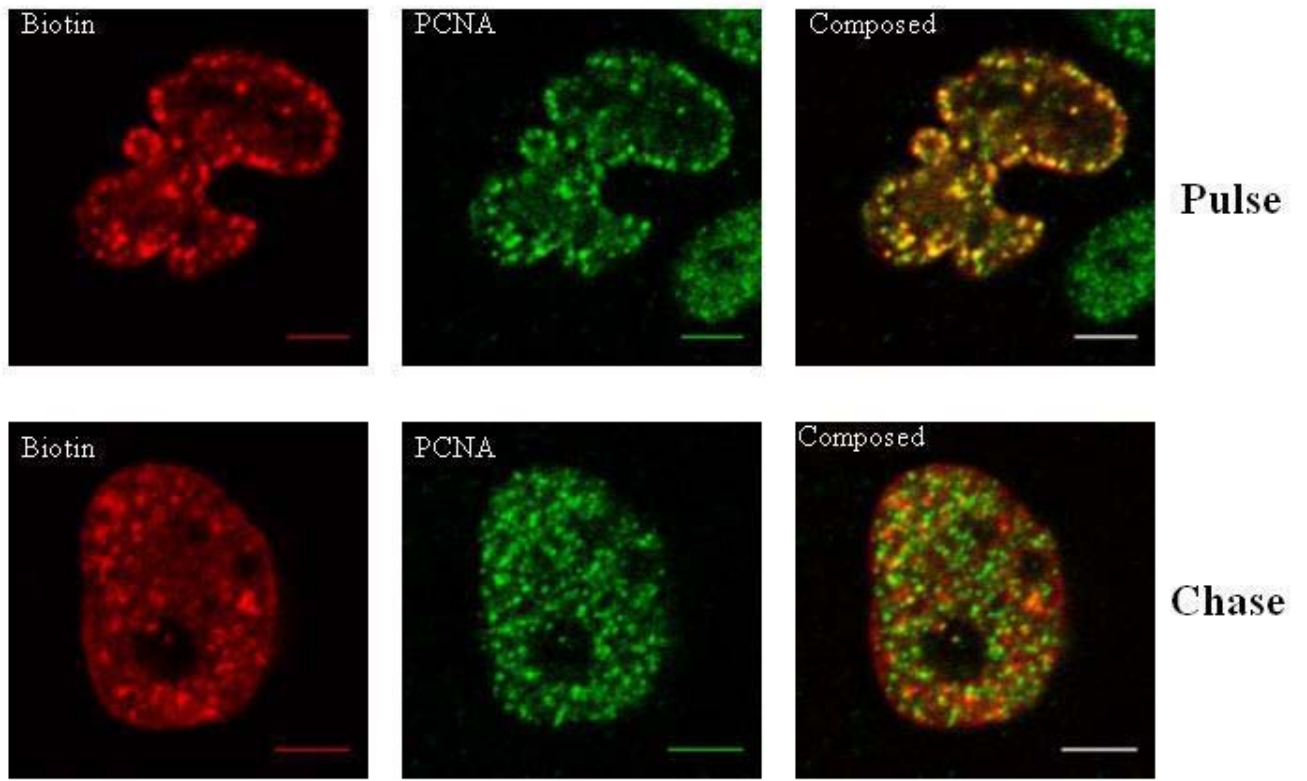

Figure S3. PCNA 6hr pulse and chase

**Figure S6**

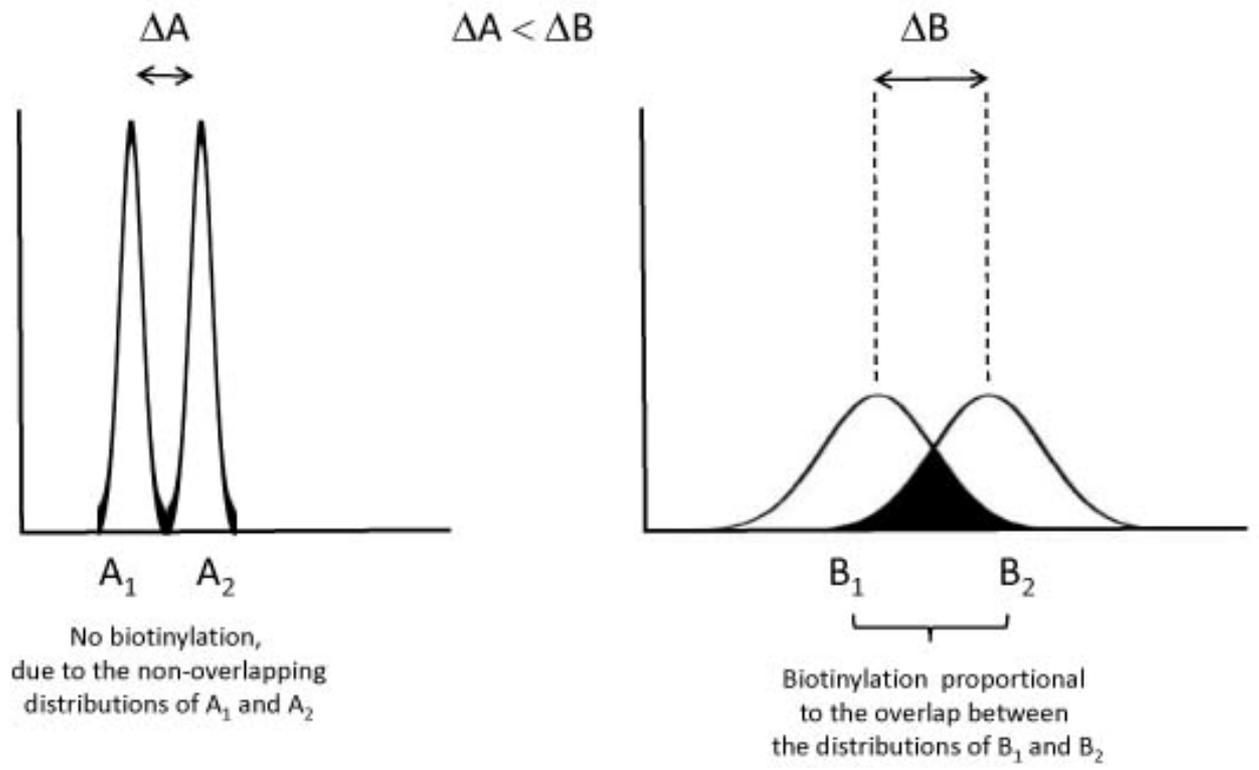

**Figure S7**